\documentclass[preprint2]{aastex6}
\usepackage{longtable}

\begin{document}

\title{A search for rapidly pulsating hot subdwarf stars in the GALEX survey}
\received{June 1, 2017}
\revised{June 28, 2017}
\accepted{June 28, 2017}

\author{Thomas M. Boudreaux\altaffilmark{1},  Brad N. Barlow\altaffilmark{1}, Scott W. Fleming\altaffilmark{2}, Alan Vasquez Soto\altaffilmark{1}, Chase Million\altaffilmark{3}, Dan E. Reichart\altaffilmark{4}, Josh B. Haislip\altaffilmark{4}, Tyler R. Linder\altaffilmark{5}, and Justin P. Moore\altaffilmark{4}}
\affiliation{\altaffilmark{1}Department of Physics, High Point University, One University Parkway, High Point, NC 27268, USA \\ \altaffilmark{2}Space Telescope Science Institute, 3700 San Martin Dr. Baltimore, MD 21218, USA\\
\altaffilmark{3}Million Concepts LLC, PO Box 119, 141 Mary St, Lemont, PA 16851, USA \\
\altaffilmark{4}Department of Physics and Astronomy, University of North Carolina, Chapel Hill, NC 27599, USA\\
\altaffilmark{5}Department of Physics, Eastern Illinois University, 600 Lincoln Ave., Charleston, IL 61920, USA
}

\begin{abstract}
NASA's Galaxy Evolution Explorer (GALEX) provided near- and far-UV observations for approximately 77 percent of the sky over a ten--year period; however, the data reduction pipeline initially only released single NUV and FUV images to the community. The recently released Python module gPhoton changes this, allowing calibrated time--series aperture photometry to be extracted easily from the raw GALEX data set. Here we use gPhoton to generate light curves for all hot subdwarf B (sdB) stars that were observed by GALEX, with the intention of identifying short--period, p-mode pulsations. We find that the spacecraft's short visit durations, uneven gaps between visits, and dither pattern make the detection of hot subdwarf pulsations difficult. Nonetheless, we detect UV variations in four previously known pulsating targets and report their UV pulsation amplitudes and frequencies. Additionally, we find that several other sdB targets not previously known to vary show promising signals in their periodograms.  Using optical follow--up photometry with the Skynet Robotic Telescope Network, we confirm $p$-mode pulsations in one of these targets, LAMOST J082517.99+113106.3, and report it as the most recent addition to the sdBV$_r$ class of variable stars.
\end{abstract}


\section{Introduction} \label{sec:intro}
Hot subdwarf B stars (sdBs) are extreme horizontal branch stars believed to have formed from red giants that lost their outer H envelopes while ascending the red giant branch, likely due to interactions with a nearby companion \citep{Heb16}. The leftover core of the progenitor star --- which becomes an sdB upon core He ignition ---has an effective temperature $22000 \leq T_{\rm{eff}} \leq 40000$ and a surface gravity $5.0 \leq \log{g} \leq 6.2$. Theory predicts sdBs should have masses around 0.5 M$_{\odot}$, which is generally consistent with reported observations \citep{Han03}.

	Subdwarf B stars are quite common, outnumbering white dwarfs down to magnitude B$\sim$18; despite this, they are one of the less well understood branches of stellar evolution. sdBs play interesting roles in our understanding of several astrophysical phenomena, including the effects of main sequence evolution interrupted by binary interactions, the UV--upturn in giant elliptical galaxies \citep{Bro97}, the ``second-parameter'' problem in globular cluster morphology (e.g., \citealt{Mon08}), and even sub-luminous Type 1a supernovae (e.g., \citealt{Gei13}). Luckily, some hot subdwarfs pulsate, and these pulsations serve as efficient probes of the interior structures and dynamics that drive this phase of stellar evolution.   
    
    The first pulsating sdB (sdBV) star, EC 14026-2647, was discoverd two decades ago by \citet{Kil97}; since then, over 100 such pulsators have been uncovered. sdBV stars come in three main flavors: (i) the sdBV$_{r}$ stars, which exhibit rapid, acoustic--mode (p--mode) oscillations with periods from 1-10 minutes and amplitudes typically $<$20 parts per thousand (ppt); (ii) the sdBV$_{s}$ stars, which exhibit slow, gravity--mode (g--mode) oscillations with periods from 1--2 hours and amplitudes around a few ppt; and (iii) the hybrid sdBV$_{rs}$ stars, which exhibit both p--mode and g--mode oscillations. Past asteroseismological studies of sdB stars, especially those using data from the {\em Kepler} mission, have led to precise measurements of sdB masses, radii, rotation rates, and other parameters (e.g., \citealt{Ost14}). The first step to unlocking the potential of asteroseismology is, of course, the discovery of new pulsating stars. Most studies of sdBV stars and searches for new pulsators have taken place in optical bandpasses, even though the sdB Planck distribution peaks in the UV and most sdBV pulsation modes have higher amplitudes in the UV compared to the optical \citep{Heb16}. 
    
    NASA's Galaxy Evolution Explorer \citep[GALEX, ][]{Mar05} provides a unique opportunity to study variable hot subdwarf stars, due to its large field coverage in UV bands.  Launched in 2003, GALEX observed 77\% of the sky through two broadband UV filters, centered around 1728 \AA\ (``FUV'') and 2271 \AA\ (``NUV''). The original data reduction pipeline yielded calibrated images of each field at a full visit depth. These, along with source catalogs, compose the primary, mission-produced archive products \citep{Mor07}. However, due to GALEX's use of micro--channel plate detectors (MCP), which recorded the individual photon events with a high degree of time accuracy, the raw GALEX data set does contain time series information. A Mikulski Archive for Space Telescopes (MAST) archive software tool called gPhoton extracts calibrated time series information on demand from the raw data by substantially reproducing key functionality from the GALEX mission calibration pipeline \citep{Mil16}.

    Here we present a search for short--period sdB pulsations in the archived GALEX dataset using gPhoton. An initial sample of 5613 hot subdwarfs \citep{Gei16}, which represents a good approximation of all cataloged hot subdwarf stars, was down-selected based on magnitudes, coordinates and total exposure time available in the gPhoton database, described fully in Section \ref{GALEX1}. These selection criteria yielded 1881 targets upon which we focused our investigation. Calibrated light curves with time bins of 30 seconds were generated for each target using gPhoton.  We identify NUV pulsations consistent with previous optical observations for four known pulsating sdB stars. Additionally, we identify several new candidate pulsators that show signals consistent with those of pulsating sdBs, and confirm one of these as a new sdBV$_r$ star with ground--based follow--up observations.
    
\section{Data Reduction with gPhoton}\label{GALEX1}
We used the gPhoton software package to produce calibrated light curves of all sdB targets. To generate light curves, the gPhoton tool called gAperture integrates sky-mapped GALEX photon events, produced by the mission with time resolutions of five microseconds, over user-defined time bins and photometric apertures, appropriately calibrated for detector exposure time and relative response \citep{Mil16}. The gPhoton package also includes a tool called gFind for quickly determining available exposure time coverage of specific targets and a tool called gMap for generating image and ``movie'' files of GALEX observations. We made use of gFind, gMap and gAperture to select targets, create 2D and 3D FITS images, and generate photometrically calibrated light curves, respectively. 
	
For each target, we extracted target ID, source position (as right ascension and declination in J2000 decimal degrees), V magnitude, and GALEX NUV magnitude when available. Note that due to both higher flux values and wider GALEX coverage in the NUV compared to the FUV, we focused our efforts on NUV measurements. Targets that fall outside of our acceptable magnitude range ($13 \leq NUV \leq 19$) are rejected. This cut conservatively eliminates bright sources that will trigger non--linear detector response and dim sources with poor signal--to--noise. All remaining targets then have their coordinates queried using gFind, which returns a data structure containing the total available exposure time, the nearest GALEX merged catalog (MCAT) source, and a breakdown of visits. Note that the MCAT is the mission-produced catalog of detected sources for all GALEX visits, but does not account for duplicate sources due to field overlaps. Also note that a ``visit'' is the amount of time spent by GALEX observing a given pointing while the spacecraft was behind earth's shadow, and can be no longer than 30 minutes in duration.

 GALEX conducted three main surveys: the All-sky Imaging Survey (AIS), Medium-imaging Survey (MIS), and Deep-imaging Survey (DIS). The All-sky Imaging Survey took $\sim$100s integrations \citep{Mor07}, too short to be useful for our investigation. Consequently we only investigate data from MIS ($\sim$1500s) and DIS ($\sim$30000s). We use the key {\lq{expt}\rq} in both NUV and FUV to select only those targets that have more than 600s in either band; this 600s cut is also used as the filter for AIS observations . After the initial 5613 sdBs provided by \citet{Gei16} were run through these criteria, we find 1881 targets with a sufficient amount of GALEX observation time to allow for pulsation searches. These targets were each visited by the spacecraft between 1 and 375 times, with a mean (median) of 7.4 (4) visits per target. The visit lengths ranged from  10--30 min, with an average visit length of about 15.5 min.
	
We used gMap to produce both full depth (coadd) images of targets using all available GALEX observations and movie files of targets with ten second integrations / frames across all available observations. A custom Python tool (FaRVaE\footnote{https://github.com/tboudreaux/FaRVaE}) was developed to automatically define the radius of the photometric aperture and the radii of the inner and outer annuli used to determine the background. The tool makes use of SEP, a software suite used to conduct aperture photometry based on Source Extractor \citep{Barb16, Ber96}. Each FITS image and cube were read into FaRVaE, where the auto-definition routine was run. We then manually verified the quality of these parameters by eye. Specifically, we ensure that there are as few bright sources in the annulus as possible, and that all visible flux is included in the aperture. We also took the opportunity to visually check the images for any obvious contamination of the detector hotspot mask into the target or for obvious astrophysical flaring activity.
	
\begin{figure}
	\centering
	\includegraphics[scale=0.33]{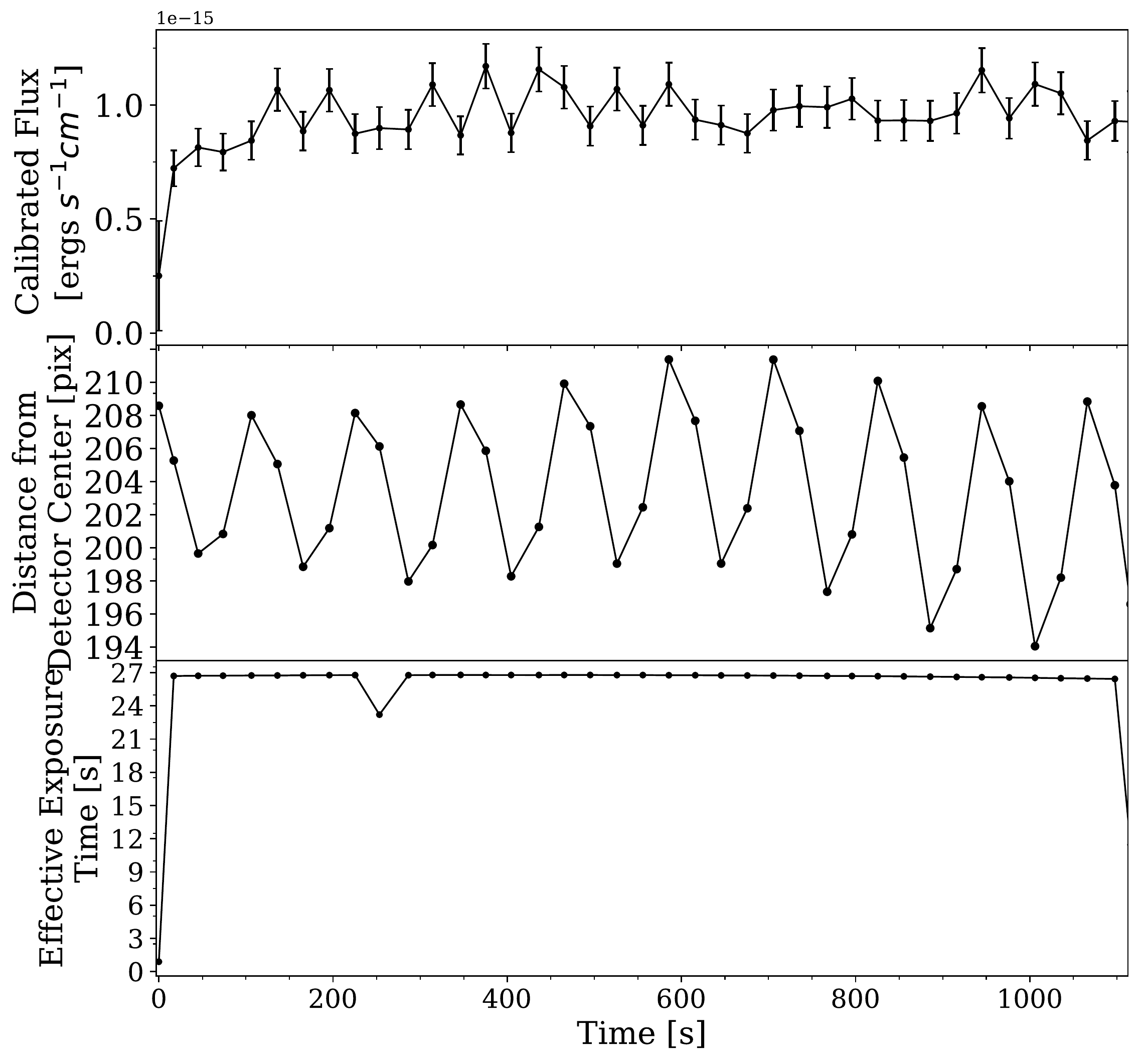}
    \caption{Example gPhoton output for one visit of one target (SDSSJ 145736.81+592927.6), including flux--calibrated light curve (top), target distance from detector center over the observation (middle), and effective exposure time (bottom).}
    \label{SumPlot}
\end{figure}

The aperture and annulus definition files were used as inputs to the gAperture module to generate aperture photometry at 30-second bins. We settled on this particular exposure time since longer cycle times would have associated Nyquist frequencies below those of some known sdB pulsations, and shorter exposure times would decrease the signal-to-noise ratio in each bin to levels that would make pulsation detection difficult, especially for low amplitudes. Due to the computationally expensive nature of a gAperture call, a consequence of network bandwidth and available computational resources, we ran the majority of gAperture calls on a cluster local to the MAST in Baltimore. Each target was run as a separate job on a 64-core machine to allow for multiple targets to run through gAperture at a time. All targets run through gAperture and gMap produced a total of 20 GBs of data, including images and light curves. Extracted output includes raw counts, calibrated fluxes, effective exposure time of each bin after accounting for dead time, the mean observation time of each bin, the mean position of the target on the detector during each bin, and associated errors. Consult the gPhoton User's Guide for a detailed description of all the available output\footnote{https://github.com/cmillion/gPhoton/blob/master/docs/\\UserGuide.md}. An example of gPhoton output for one of our targets is shown in Figure \ref{SumPlot}.    

\section{Data Analysis}\label{GALEX2}
\begin{figure}
	\centering
    \includegraphics[scale=0.33]{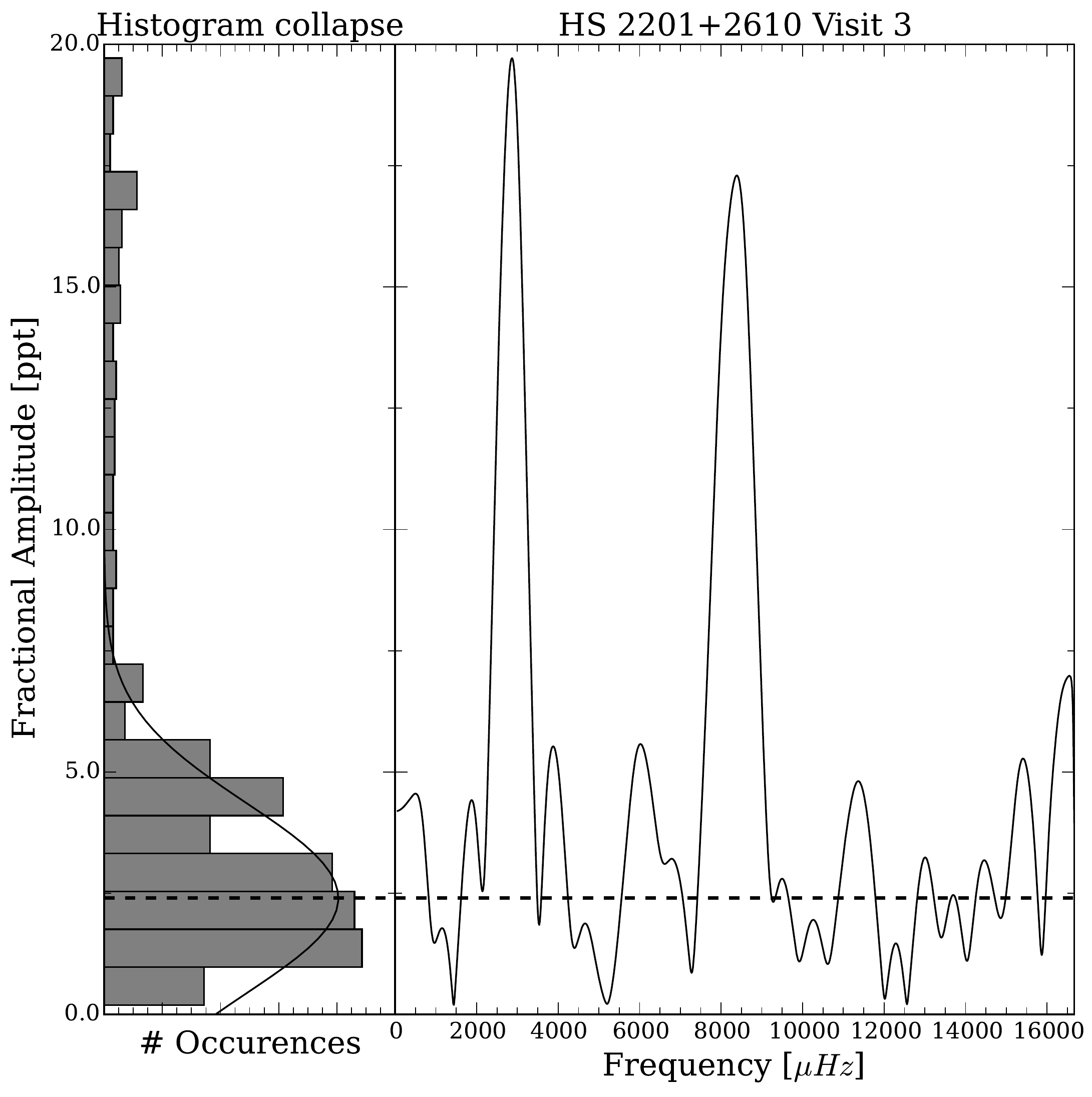}
    \includegraphics[scale=0.33]{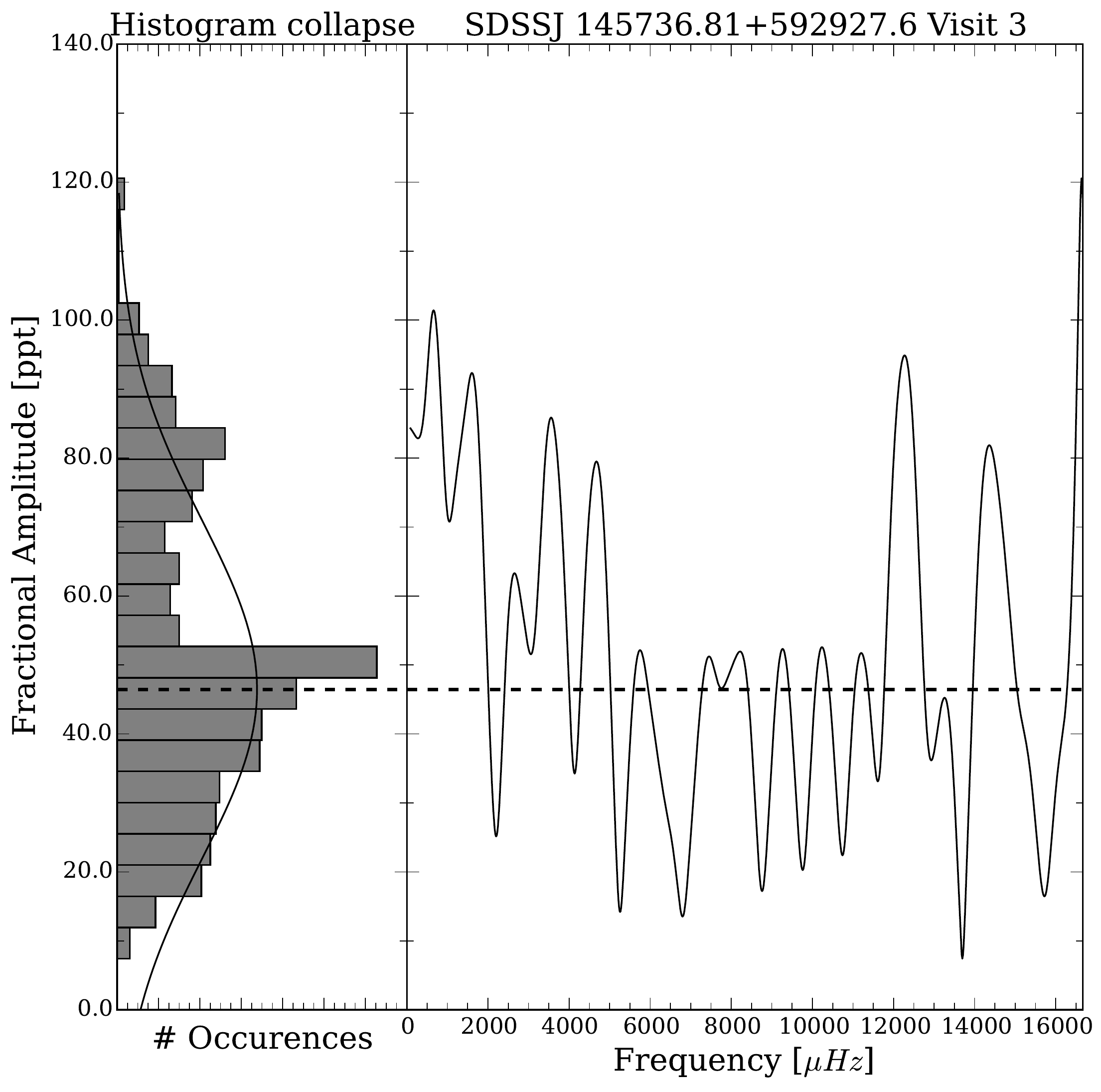}
    \caption{Lomb-Scargle periodiograms (right panels) and their projections on the amplitude axis (left panels) for example targets HS 2201+2610 Visit 3 (Top) and SDSSJ 145736.81+592927.6 Visit 3 (Bottom). The approximate mean noise level in each LSP (dashed line) is calculated from a Gaussian fit to the amplitude histogram plot.}
    \label{FTCollapse}
\end{figure}

Given the large number of light curves generated by gPhoton (13919 in total), we decided not to look at each individual light curve by eye for photometric variations. Moreover, sdBV amplitudes tend to be small (1-30 ppt) and easily hidden by noise, generally requiring a Fourier transform for identification and analysis.  We compute the Lomb-Scargle periodogram (LSP) \citep{Lom76, Sca82} --  as implemented by the SciPy library \citep{Oli07,Mil11} -- for each individual light curve in order to look for periodicities and determine their frequencies and amplitudes. As GALEX observed over a ten-year timespan (2003-2013), much of the data returned from gPhoton for a particular target has large gaps between spacecraft visits, in excess of a year in some cases. We decided to avoid problems associated with welding together and analyzing data with such large gaps in between, and instead analyze the light curves for each target on a visit--by--visit basis.  Example LSPs for two of our targets are shown in the right panels of Figure \ref{FTCollapse}.

\begin{figure*}[t]
	\centering
	\includegraphics[scale=0.3]{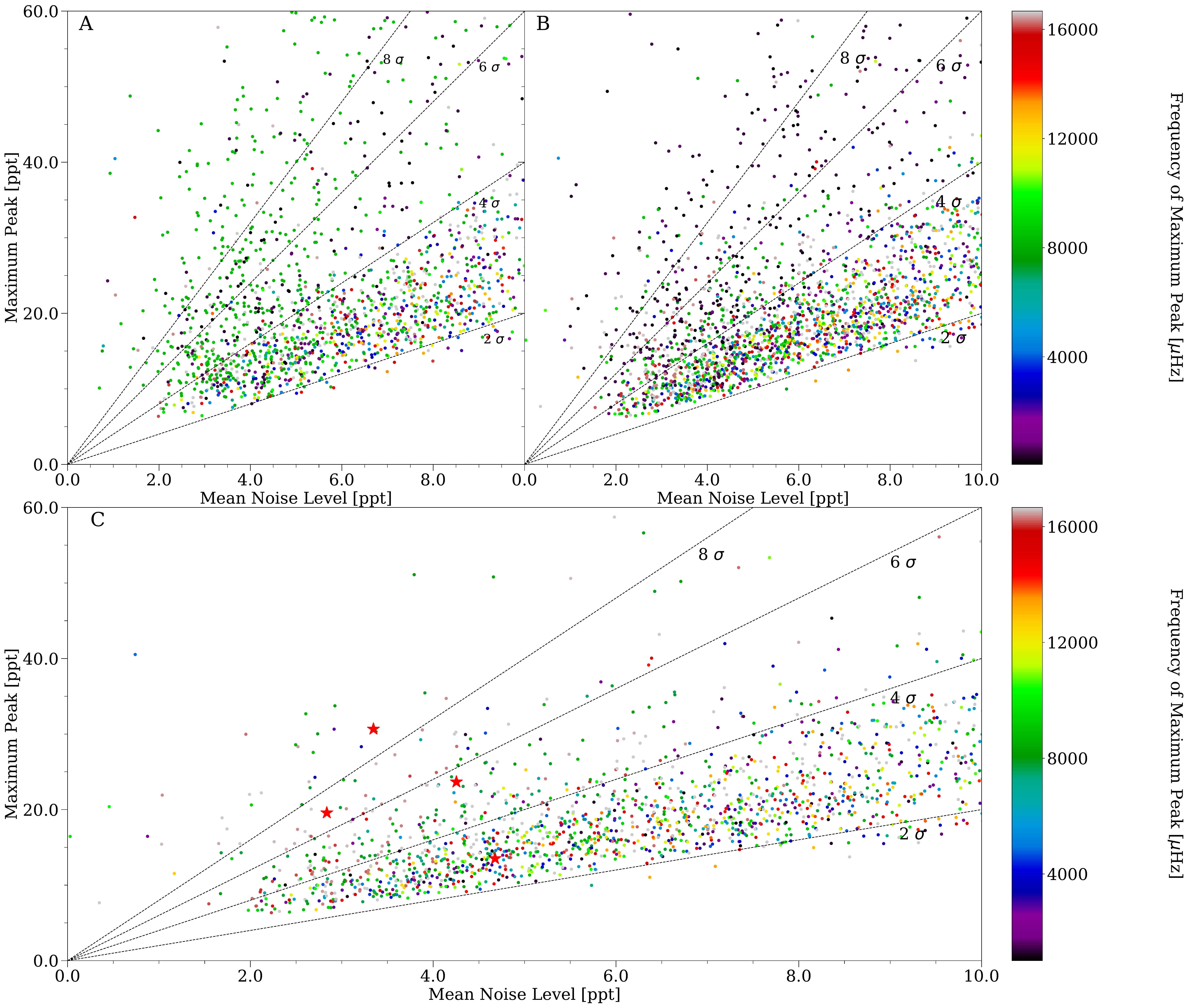}
    \caption{Maximum Peak in the Lomb-Scargle Periodogram (LSP) vs. Mean Noise in the LSP. (a) - No pre-whitening, all visits plotted, (b) - pre-whitening, all visits plotted, (c) - pre-whitening, visits with a maximum peak amplitude lower than 1000 $\mu Hz$ not plotted. Star symbols mark identified known pulsating sdBs -- from left to right, HS 2201+2610 \citep{Ost01}, GALEX J0869+1527\citep{Bar11}, EC 14026-2647 \citep{Kil97}, HS 0815+4243 \citep{Ost01}.}
    \label{MPFFinal}
\end{figure*}

Candidate pulsators can be identified by comparing the highest peak in each LSP to its corresponding mean noise level $\sigma$. While maximum peak values are simple to extract from the periodograms, mean noise levels prove to be more difficult to estimate given the short duration of each visit. Initially, the RMS scatter about the mean for each visit's light curve was used as a mean noise level estimate -- however, these values were consistently high relative to the apparent noise level (by visual inspection) in the LSP. The poor frequency resolution in the single--visit LSPs, around $\sim$667 $\mu$Hz, permits strong signals (whether real or not) to raise the estimated noise level above its actual value, thereby making the signals appear at lower S/N than they are. We settled on what we found to be a relatively robust method:  we ``collapse'' the power spectrum onto the amplitude axis, plot a histogram of amplitude values, and fit a standard Gaussian function to this distribution.  We take the centroid of this Gaussian fit as the mean noise level for the LSP.  As illustrated in the left panels of Figure \ref{FTCollapse}, this method keeps noise spikes and actual stellar variations from skewing the estimated noise level, thereby permitting us to use the S/N of the highest peak to assess its significance properly.

	Summarizing our entire data set, we plot in Figure \ref{MPFFinal}a the maximum peak amplitude in each LSP against the mean noise level. Additionally, we color each point according to the frequency associated with the highest LSP peak. Target visits with stronger photometric variations will appear at larger angles off the positive horizontal axis (at higher $\sigma$ values). The vast majority of points fall between 2$\sigma$ and 4$\sigma$, indicating no significant variations above the noise level. 
    
\begin{figure}
	\centering
     	\includegraphics[scale=0.33]{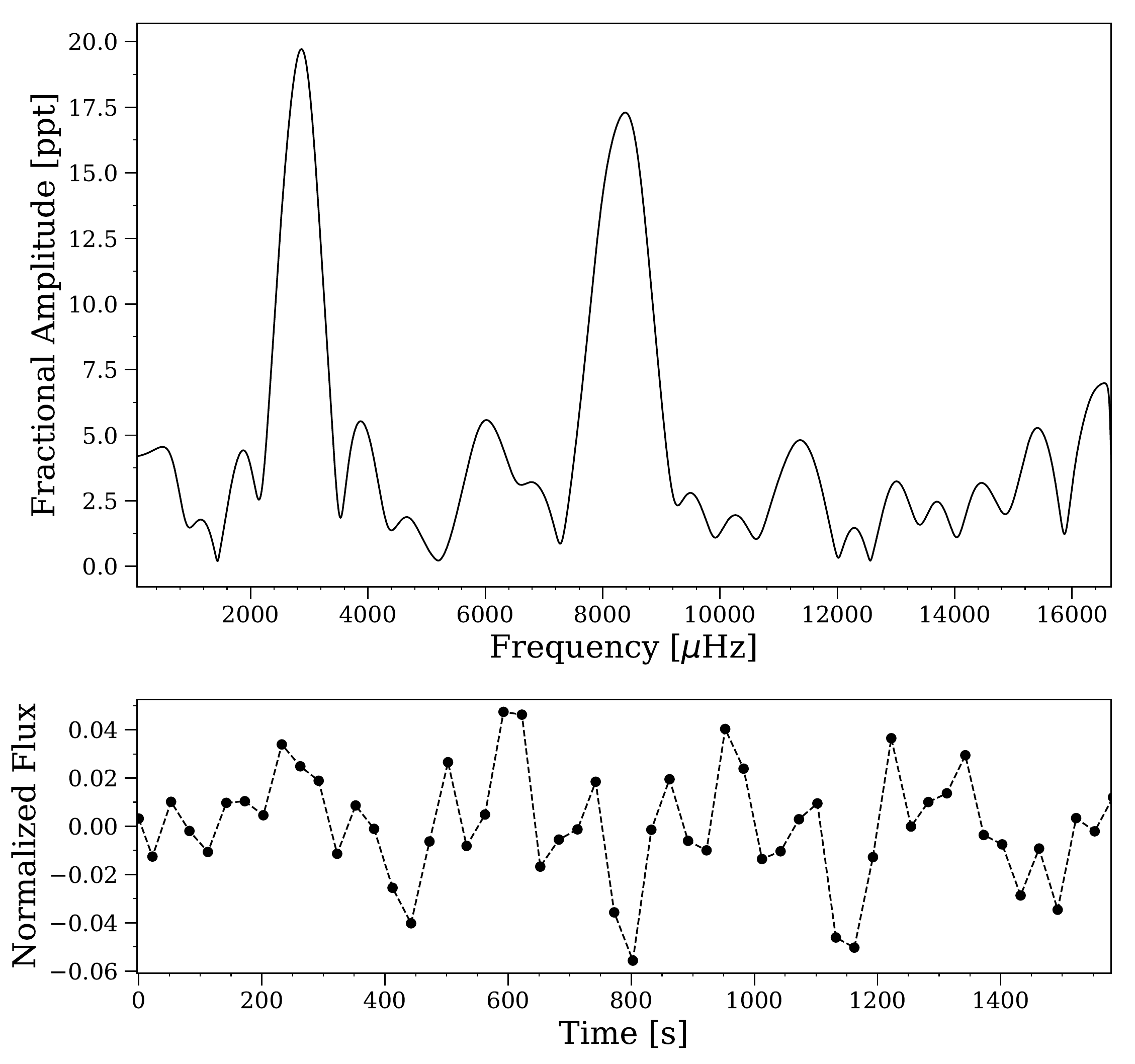}
        \includegraphics[scale=0.33]{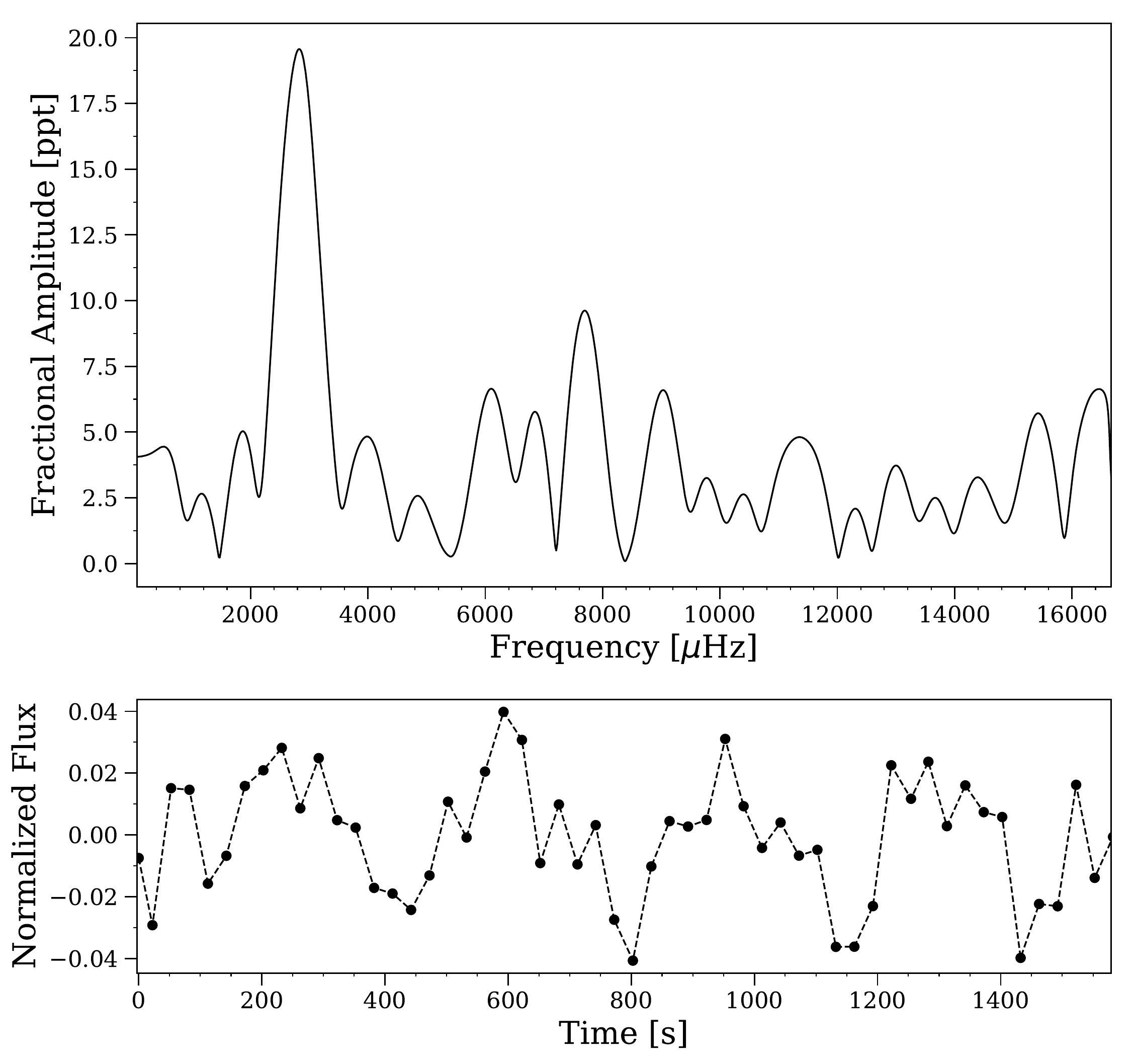}
        \caption{HS 2201+2610 light curve and LSP before (top) and after (bottom) pre-whitening the dither pattern alias (at $\approx8000\mu Hz$). We note that other features are not significantly affected by the pre-whitening, most importantly the stellar pulsation near 2800 $\mu$Hz.}
        \label{BAPW}
\end{figure}

    Careful observation of Figure \ref{MPFFinal}a reveals a predominance of visits with strong signals around 8300 $\mu$Hz ($\sim$120 s; green points) -- a phenomenon which was originally not expected. Investigating light curves exhibiting this signal by eye reveals a clear correlation between flux and position on the detector (``detrad'').  The detrad variation and its frequency are consistent with the so--called ``petal'' dither pattern of the GALEX spacecraft.  In some cases, we found that this pattern even generates a false signal at its first harmonic, out near 16000 $\mu$Hz. Consequently, we decided to pre-whiten all light curves of this instrumental artifact.  First, we fit the sum of two sine waves to each light curve, one with frequency fixed to 8341 $\mu$Hz and amplitude fixed to the amplitude of this signal in the LSP, and another with frequency and amplitude fixed to those of the first harmonic of the dither pattern. The best-fitting sine waves are then subtracted from each light curve to remove the petal pattern, and new LSPs are calculated. Figure \ref{BAPW} shows the light curve and LSP for one of our target visits, before and after the pre-whitening of the petal pattern signal.
    	
    The newly pre-whitened target light curves are run through the same scripts previously discussed to generate Figure \ref{MPFFinal}b. The predominance of points around 8000 $\mu$Hz (green points) is now gone. We investigated the large number of targets remaining with maximum peak frequencies below 1000 $\mu$Hz (dark purple/black points) -- a regime where the LSP is dominated by 1/f noise -- and find that many of these visits have a long-term variation introduced by a second, lower frequency spacecraft dither pattern. We elected to remove this frequency range from the calculation of the highest LSP peak for three reasons:  (i) signals in this range are likely due to 1/f noise or a known, longer--period spacecraft dither pattern; (ii) these low--frequency signals can overpower true signals at other frequencies; and (iii) sdBV$_r$ pulsations are not expected at frequencies lower than 1000 $\mu$Hz anyway, so the likelihood of missing stellar pulsations at f $<$ 1000 $\mu$Hz is low. After this low frequency cut the most visits a target has is 119 with a mean (median) of 7.4 (2) visits per target.
    
    Figure \ref{MPFFinal}c summarizes our full dataset after removing or ignoring instrumental effects and 1/f noise. We present in Table \ref{tab:TBPW} a small subset of target measurements used to produce Figure \ref{MPFFinal}c, with the entire set available electronically. The plot bounds of Figure \ref{MPFFinal} were chosen in order to highlight the region where sdBVr pulsations are expected to exist. One will notice a sharp drop-off in point density above the 4$\sigma$ line, compared to panels a and b (in which the dither pattern and 1/f noise dominate many LSPs). In essence, Figure \ref{MPFFinal}c provides an ordered list of targets to investigate for stellar pulsations, starting with the highest S/N objects.  Before using these data to search for new pulsators, however, we attempted to recover NUV signals from {\em known} pulsating sdBV$_r$ stars. 

\begin{table*}
\centering
\begin{tabular}{l c c c c c c c}
\hline
\hline
                      &  Visit &    Start Time &  Visit Length &   Mean Noise &  Max Peak &   Frequency &       Sigma \\
                     Target ID & [\#] & [MJD] & [s] & [ppt] & [ppt] & [$\mu$Hz] \\ 
\hline
                   PG 0039+049 &      1 &  54721.2825925 &      1059.970 &     1.6 &  285.5 &    704 &  183.9 \\
                               &      2 &  54747.1058281 &       802.797 &     39.0 &  310.6 &    567 &    8.0 \\
                  FBS 2227+383 &      1 &  55020.5734781 &      1129.439 &     10.3 &  541.2 &    809 &   52.6 \\
                               &      2 & 55058.98114985 &       860.677 &     33.4 &  416.4 &   1057 &   12.5 \\
                   PG 1716+426 &      1 &  55049.3895511 &      1644.966 &     3.1 &  111.5 &    616 &   36.4 \\
                       PB 7409 &      1 &  55081.7076462 &      1643.916 &     15.5 &  465.1 &    561 &   30.1 \\
                               &      2 &  55108.2046817 &      1575.557 &     41.0 &  473.4 &    563 &   11.5 \\
\hline
\end{tabular}\caption{Sample results from our data analysis, showing seven GALEX visits to four sdBs with over 600s of exposure time. Mean noise, maximum peak, and frequency of maximum peak are all reported after pre-whitening for the dither pattern and the first harmonic of the dither pattern. Start Time refers to the beginning of the GALEX visit, where MJD = JD - 2400000.5}. The entirety of this table is available electronically.
\label{tab:TBPW}
\end{table*}

\section{Detections of Known sdBV$_r$ stars} \label{OTFP}
\begin{figure*}
	\centering
	\begin{tabular}{c c c}
    	\includegraphics[width=0.33\textwidth]{fig6} & \includegraphics[width=0.33\textwidth]{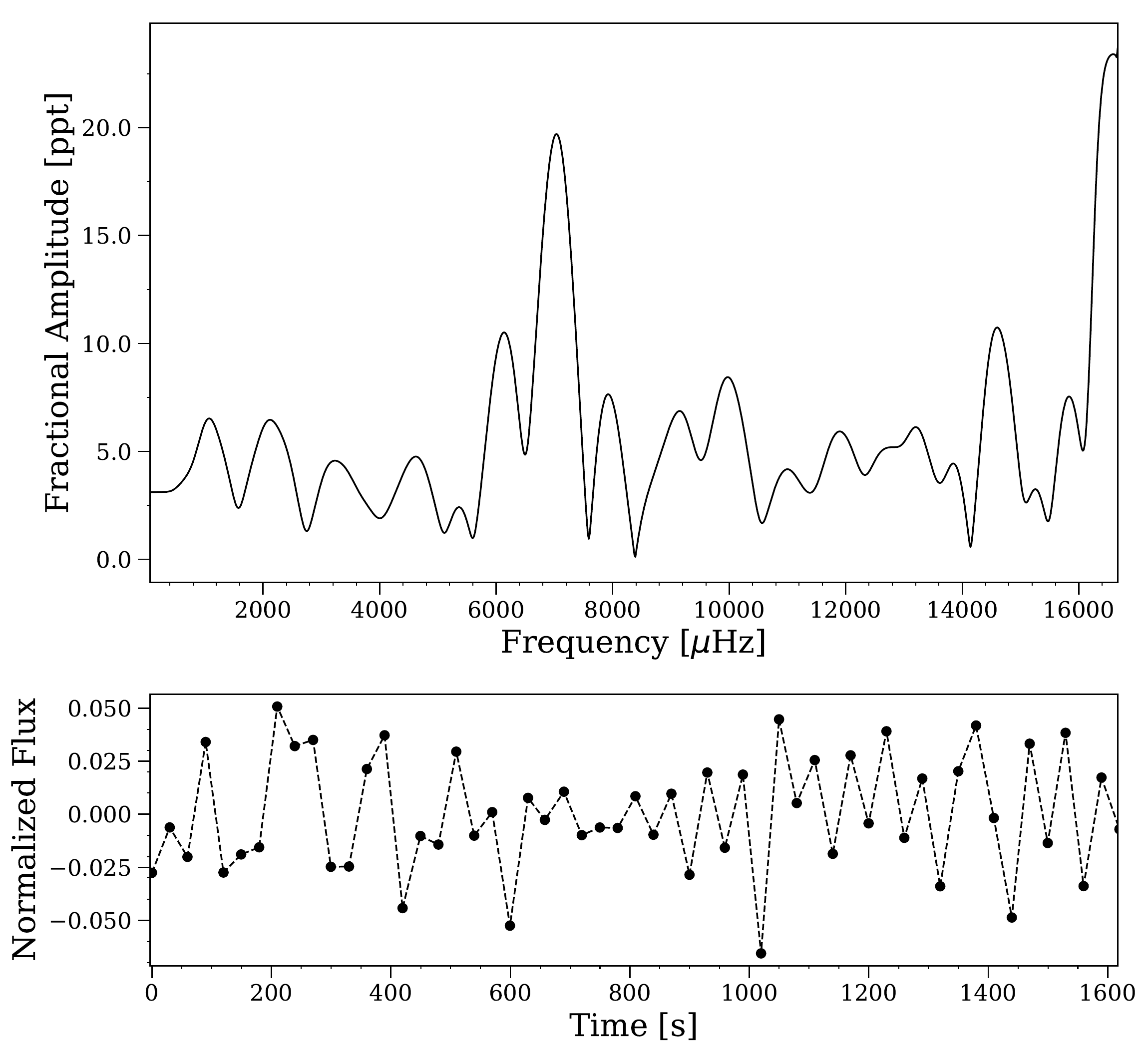} & \includegraphics[width=0.33\textwidth]{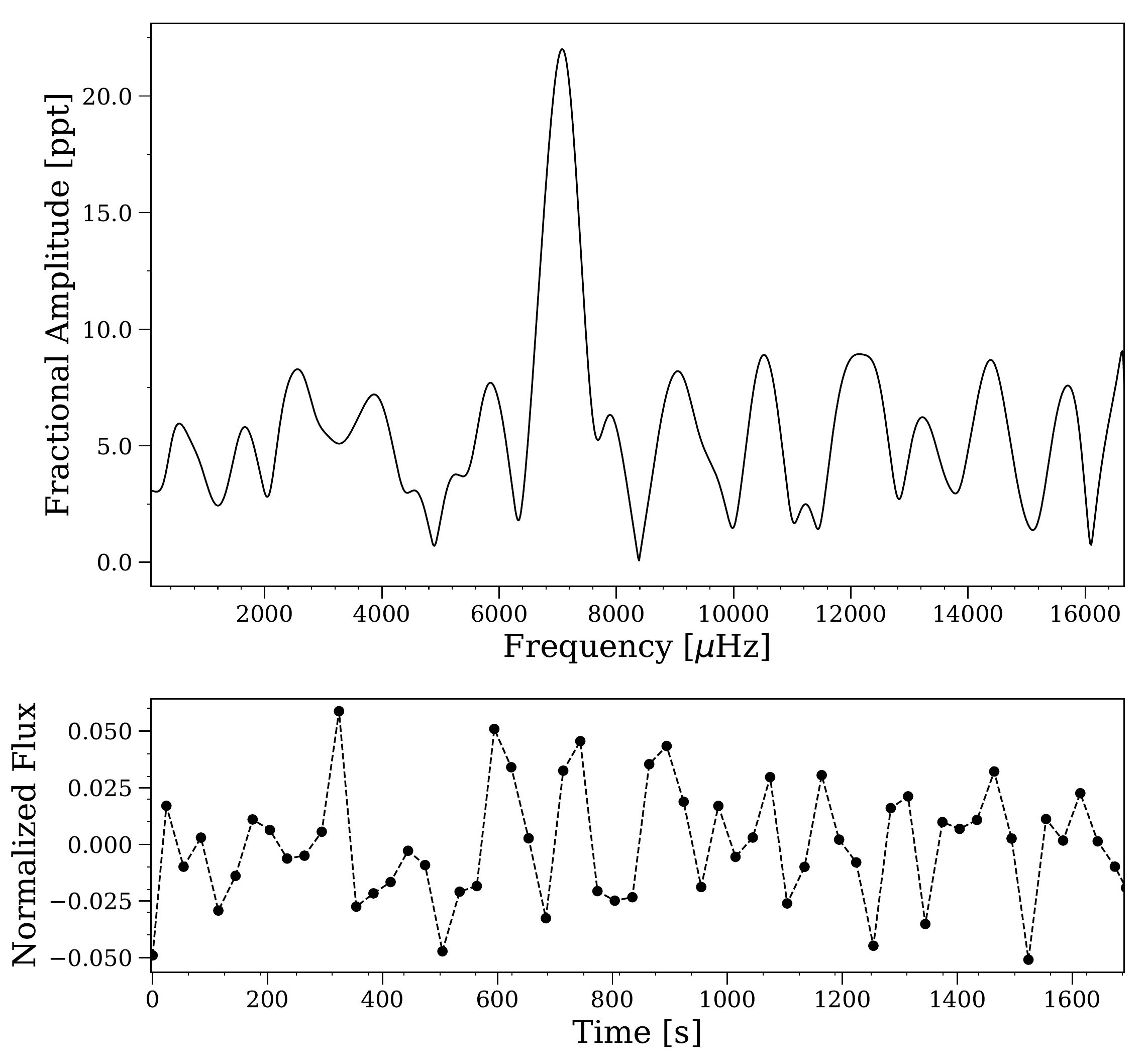} 
    \end{tabular}
    \begin{tabular}{c c}
    	\includegraphics[width=0.33\textwidth]{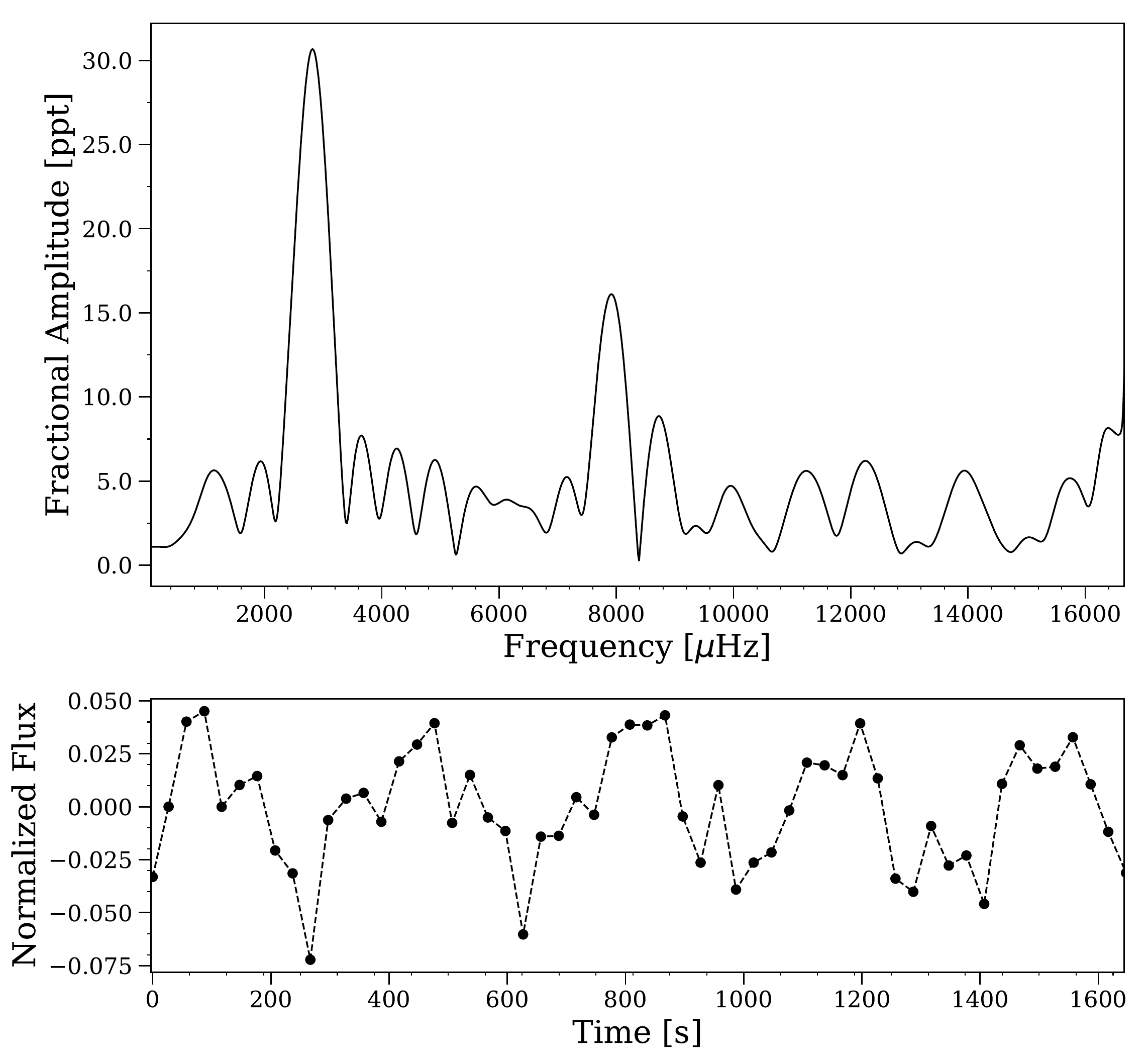} & \includegraphics[width=0.33\textwidth]{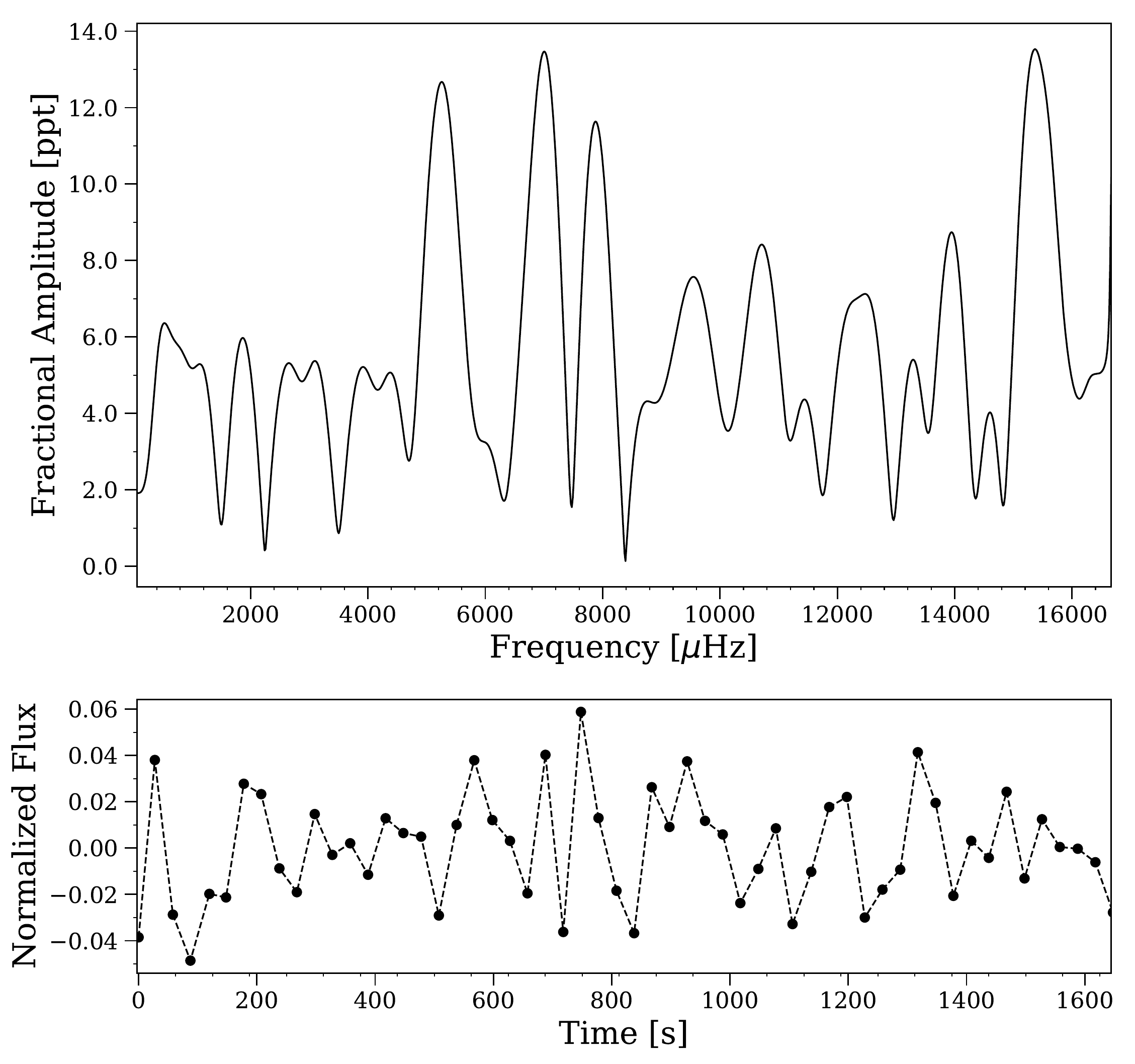}\\
    \end{tabular}
    \caption{Single--visit light curves and Lomb-Scargle periodograms for previously--known sdBV$_r$ stars detected in the GALEX NUV data set. Clockwise from top--left, targets shown include HS 2201+2610, EC 14026-2647 Visit One, EC 14026-2647 Visit Two, GALEX J08069+1527, HS 0815+4243.}
    \label{FoundLC}
\end{figure*}

	 We cross examined all known sdBV$_r$ stars found in the literature (e.g., \citealt{Ost10,Gei16}) with our data set. We find that of the thirteen known sdBV$_r$ stars with sufficient GALEX observations for analysis, shown in Table \ref{KnownTab}, we were only able to recover pulsations in four of these objects. Their NUV light curves and corresponding LSPs are shown in Figure \ref{FoundLC}. We use non-linear, least squares fitting of sine waves to the data to determine pulsation amplitudes and frequencies, which are shown in Table \ref{Known}. The wavelength dependence on a pulsation mode's amplitude (especially UV--optical comparisons) has been used in the past to identify the mode's degree index $l$, among other parameters \citep{Ran05}.  However, sdB pulsation amplitudes are known to be unstable over timescales on the order of days to years \citep{Kil10}. Without having contemporaneous optical observations (which we are not able to find for any of the four previously known sdBV$_r$ targets identified here), we do not attempt to draw any conclusions based on the comparison between our measured GALEX amplitudes to optical amplitudes. Instead, we simply report NUV amplitudes and frequencies, and their consistency with previous ground--based studies. We assess the significance of expected peaks using the regularized incomplete beta function, which describes the distribution of powers in the LSP, after normalization by the sample variance \citep{Sch98}. Brief comments on the four known sdBV$_r$ stars detected are given in the sections that follow. 
    
\begin{table*}
\centering
\begin{tabular}{l  c c c}
\hline
\hline
                      &    Start Time &  Visit Length &   Mean Noise \\
                     Target ID & [MJD] & [s] & [ppt] \\
\hline
       PG 0911+456 &   53381.979664 &      1673.683 &    2.9  \\
      HS 2201+2610* &    55829.740088 &      1582.440 &    2.8  \\
       PG 1657+416 &     52861.746673 &      1433.522 &    6.0  \\
       PG 1047+003 &     53092.802616 &      1669.538 &   71.5  \\
      HS 1824+5745 &       55820.70283 &      1558.150 &    7.2  \\
      HS 0815+4243* &      55211.044285 &      1647.533 &    4.8  \\
      HS 0039+4302 &     53683.109493 &      1672.339 &    4.6  \\
     EC 14026-2647* &     53857.464188 &	1693.600 &	4.9  \\
 \scriptsize{GALEX J08069+1527}* &      55203.24052 &      1647.039 &    3.0  \\
      HE 2151-1001 &    54679.310077 &      1070.767 &    5.3  \\
       PG 1219+533 &     55633.879924 &      1669.925 &   11.9  \\
       PG 1618+562 & 53493.344468	& 1005.4 & 	2.8 \\
       HS 2125+1105 &	55021.595772 &	887.498 &	23.2 \\
\hline
\end{tabular}\caption{Single GALEX visit for the each of the 13 known sdBV$_{r}$ targets present in our dataset. Note that the noise levels for these targets are near to or larger than the characteristic pulsation amplitude of an sdBV$_{r}$. Those targets that were identified have pulsations amplitudes greater than the norm. \\  
**  sdBV$_{r}$ identified in this study.}
\label{KnownTab}
\end{table*}

\begin{table*} 
	\centering
	\begin{tabular}{c  c  c  c }
    \hline
    \hline
     & NUV Amplitude & Frequency & Period \\
     Target ID & [ppt] & [$\mu$Hz] & [s] \\
    
    \hline
    HS 2201+2610  & 20 $\pm$ 2 & 2800 $\pm$ 45 & 357.14 \\
    EC 14026-2647 (Visit 1) & 19 $\pm$ 4 & 7030 $\pm$ 75 & 142.24 \\
    EC 14026-2647 (Visit 2) & 22 $\pm$ 4 & 7074 $\pm$ 53 & 141.36 \\
    
    GALEX J08069+1527 & 31 $\pm$ 3 & 2810 $\pm$ 32 & 355.87 \\
    HS 0815+4243 & 12 $\pm$ 4 & 7880 $\pm$ 121 & 126.9 \\
    \hline
\end{tabular}
    
    \caption{NUV amplitudes, frequencies, and associated uncertainties for known pulsating sdBV$_r$ stars with GALEX--detected NUV variations.}
    \label{Known}
\end{table*}
    
    \subsection{HS 2201+2610}
    From only one usable GALEX visit, we detect a single oscillation with frequency of 2800 $\pm$ 45 $\mu$Hz and NUV amplitude of 20 $\pm$ 2 ppt. The optical counterpart to this signal is difficult to identify, as observations by \citet{Ost01} and \citet{Sil02} show that HS 2201+2610 exhibits several signals near this period with frequency separations smaller than the resolution of our periodogram. The three largest optical signals occur at frequencies 2860, 2824, and 2880 $\mu$Hz, with $B$-filter amplitudes near 10, 4, and 1 ppt, respectively.  Our detected signal is likely a blend of these, a result of our poor frequency resolution. Nonetheless, it is clear that the pulsation amplitudes in the NUV are approximately twice as high as they are in the optical. Assuming we expect a signal near 2800 $\mu$Hz, we calculate a $2.1\times10^{-6}$ probability that a peak as large as the one observed (power $\sim$13.1) would occur there by chance.

	\subsection{EC 14026-2647}
    The prototype pulsating sdBV star, EC 14026-2647, was originally found to be dominated by a single variation of $\sim$12 ppt with a frequency around 6930 $\mu$Hz \citep{Kil97}.  On some nights, however, a second pulsation mode at 7462 $\mu$Hz was detected with an amplitude around 4 ppt.  We find in the GALEX data a single signal at 7030 $\pm$ 75 $\mu$Hz with NUV amplitude of 19 $\pm$ 4 ppt, consistent with the first of the two Kilkenny signals. With a power of $\sim$9.3 in the sample variance normalized LSP, this signal has a $9.1\times10^{-5}$ probability of occurring by chance. Similar to HS 2201+2610, the NUV amplitude for this pulsation is nearly twice as large as in the optical. We do not detect the second frequency in our data; whether this is due to a relatively poor noise level, bad frequency resolution, or the pulsation mode simply not being present at the time of the observation is unclear. 
    
   \subsection{GALEX J08069+1527}
   GALEX J08069+1527 had one useful GALEX visit, from which we report a single signal at 2810 $\pm$ 32 $\mu$Hz with amplitude 31 $\pm$ 3 ppt. This is a clear detection of the dominant pulsation mode reported by \citet{Bar11}, which had a $B$-filter amplitude of 27 ppt. The probability this peak (power $\sim$14.4) is due to noise alone is $5.6\times10^{-7}$. We do not detect the second mode reported in the optical discovery data, which would have a predicted NUV amplitude below our noise level.
   
   \subsection{HS 0815+4243}
   \citet{Ost01} reported a signal between 5.9 and 7.8 ppt (variation over the course of three observations) at a frequency of 7920 $\mu$Hz in HS 0815+4243. Using the single visit available for this target, we find four peaks in the LSP that stand out above the noise level. While the sigma value for this target is apparently quite low (especially compared to the other known pulsators identified here), it is the number of similarly large peaks -- which we believe to be predominately due to noise --  that serves to inflate the mean noise level thus deflating the sigma value. Consequently we see that we cannot rely solely on the sigma metric as it is subject to under estimation when the number of similarly large peaks is high. Instead this target is identified only by using prior knowledge of the pulsation. One of the large peaks, with amplitude 12 $\pm$ 4 ppt and frequency 7880 $\pm$ 121 $\mu$Hz, is consistent with the \citet{Ost01} detection. We find that this peak has a power of $\sim$3.03, from which we calculate a 4.8\% probability it could occur by chance. 
   
   For the other three peaks, the probability calculations are not as straightforward since we have no prior expectations for power at these frequencies. Alternatively, we use $10^6$ Monte-Carlo trials to quantify their false alarm probabilities, or the odds of a peak so high occurring somewhere between 0 Hz and the Nyquist frequency by chance. For each trial, we construct a light curve with the same observation times as the GALEX light curve and inject into it Gaussian noise with variance matching that of the observed data set. We compute the LSP of each synthetic light curve and record its maximum power. The false alarm probability of an observed signal without prior detection is equal to the fraction of trials in which the maximum power exceeds that of the observed peak. Our Monte Carlo simulations show that the three peaks at 6993, 5238, and 15368 $\mu$Hz have false alarm probabilities of $\sim$30\%, $\sim$75\%, and $\sim$45\%, respectively. Consequently, we hesitate to claim them as new detections.
    
\section{New Candidate Pulsating sdBs} 
As previously mentioned, Figure \ref{MPFFinal}c and Table \ref{tab:TBPW} effectively provide an ordered list of targets to follow--up for confirmation of stellar pulsations. Most targets in this figure fall below the 4$\sigma$ line, indicating either stellar pulsations swamped by the noise level, or the lack of pulsations altogether. Some targets, however, do show signals at higher signal--to--noise ratios. We note that many of these signals appear to be instrumental in nature, remnants of poor dither-pattern subtraction (green points -- petal pattern fundamental oscillation;  dark red points -- petal pattern first harmonic). Nonetheless, a few viable targets remain at or above the 4$\sigma$ line and warrant follow--up observations for confirmation. While a complete follow--up survey of these targets is beyond the scope of this paper, we were able to obtain sufficient ground--based observations of one candidate pulsator,  LAMOST J082517.99+113106.3 (SDSS J082517.99+113106.2), which we discuss in detail in the following section.

\subsection{LAMOST J082517.99+113106.3 --- A New sdBV}

\begin{figure}
  \centering
  \includegraphics[scale=0.33]{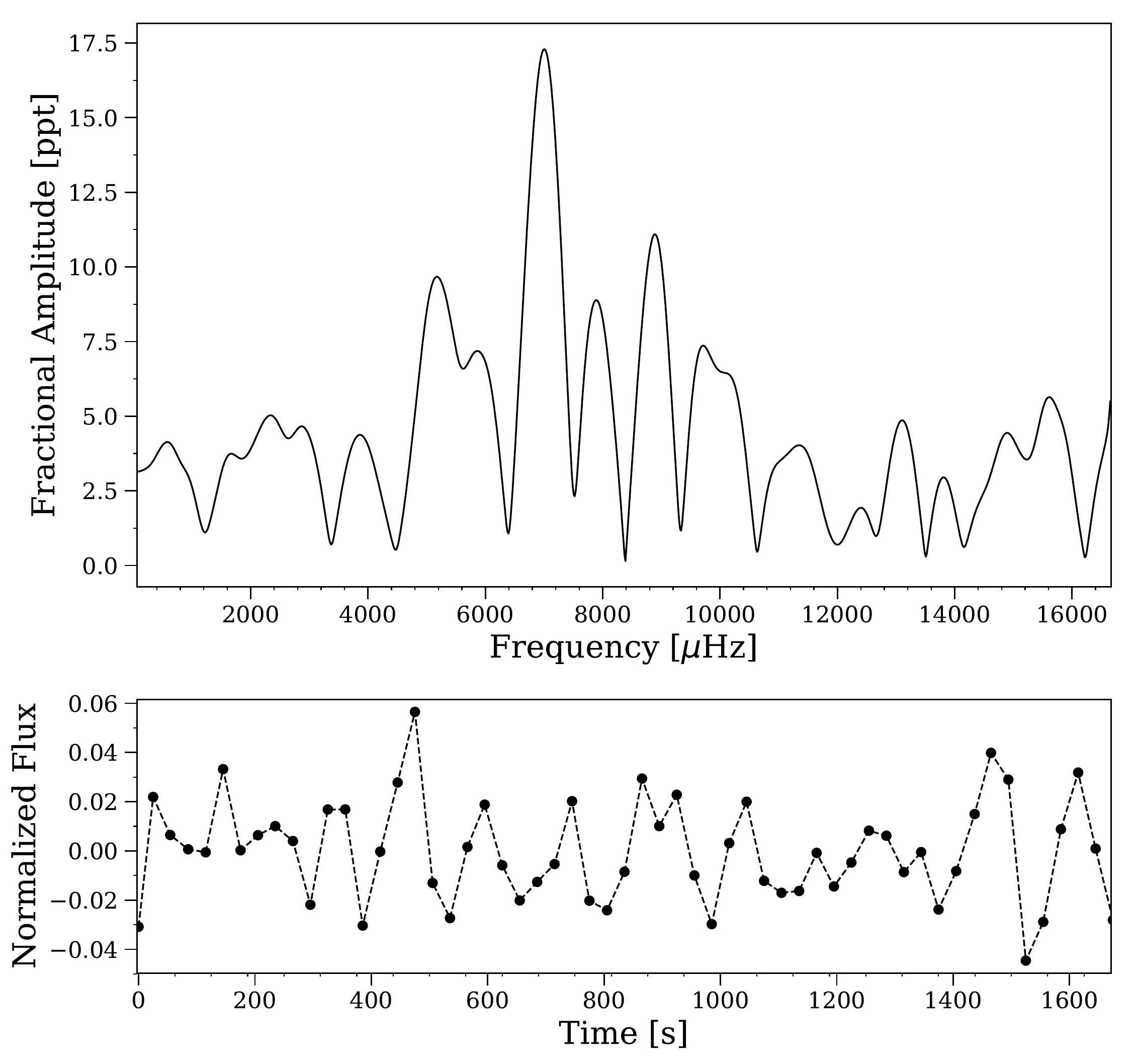}
  \includegraphics[scale=0.33]{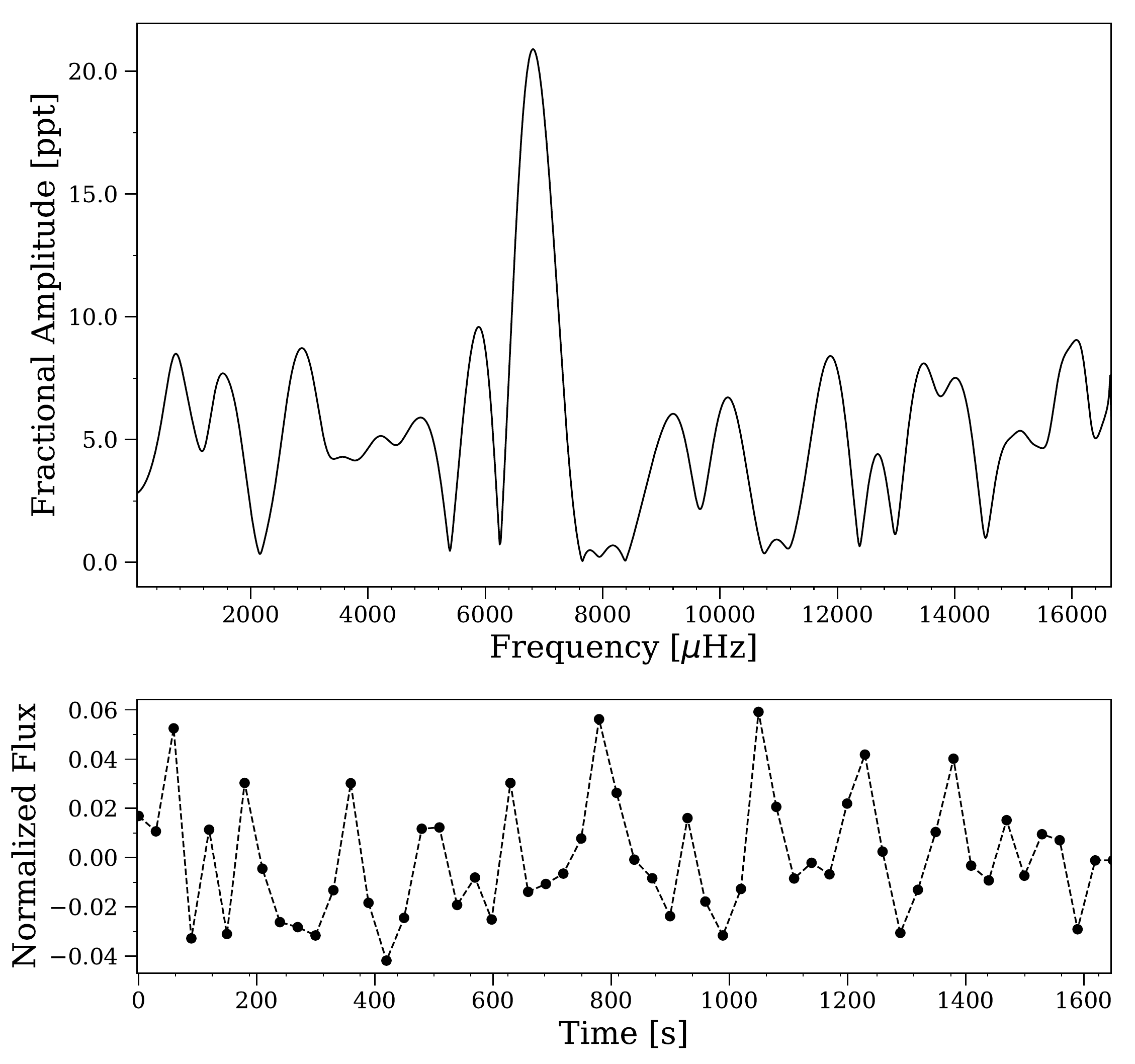}
    \caption{NUV light curve and corresponding Lomb-Scargle periodogram for LAMOST J082517.99+113106.3 Visit One (top) and Visit Two (bottom), a new candidate sdBV$_r$ star identified from the GALEX dataset.}
    \label{Lamost_GALEX}

\end{figure}

Usable GALEX data for LAMOST J082517.99+113106.3 consists of two separate visits (Figure \ref{Lamost_GALEX}). Both visits reveal the same candidate signal, which has an average NUV amplitude of 19 $\pm$ 3 ppt and frequency of 6900 $\pm$ 40 $\mu$Hz ($\sim$145 s). In order to confirm pulsations in this target, we conducted ground--based, follow--up observations on March 12, 2017 with the Skynet robotic telescope array \citep{Rei05}. We used the 0.61--m PROMPT-3 telescope, located at the Cerro Tololo Inter-American Observatory in Chile, to obtain 400 continuous images over a two-hour timespan. By using a high--throughput ``Clear'' filter, we were able to maximize the signal--to--noise ratio so that we could easily confirm the GALEX--detected pulsation mode and look for other smaller modes that might be present. Each image had an exposure time of 20 s and cycle time of 27 s, resulting in a duty cycle near 74\%. 

\begin{figure}
	\centering
	\includegraphics[scale=0.5]{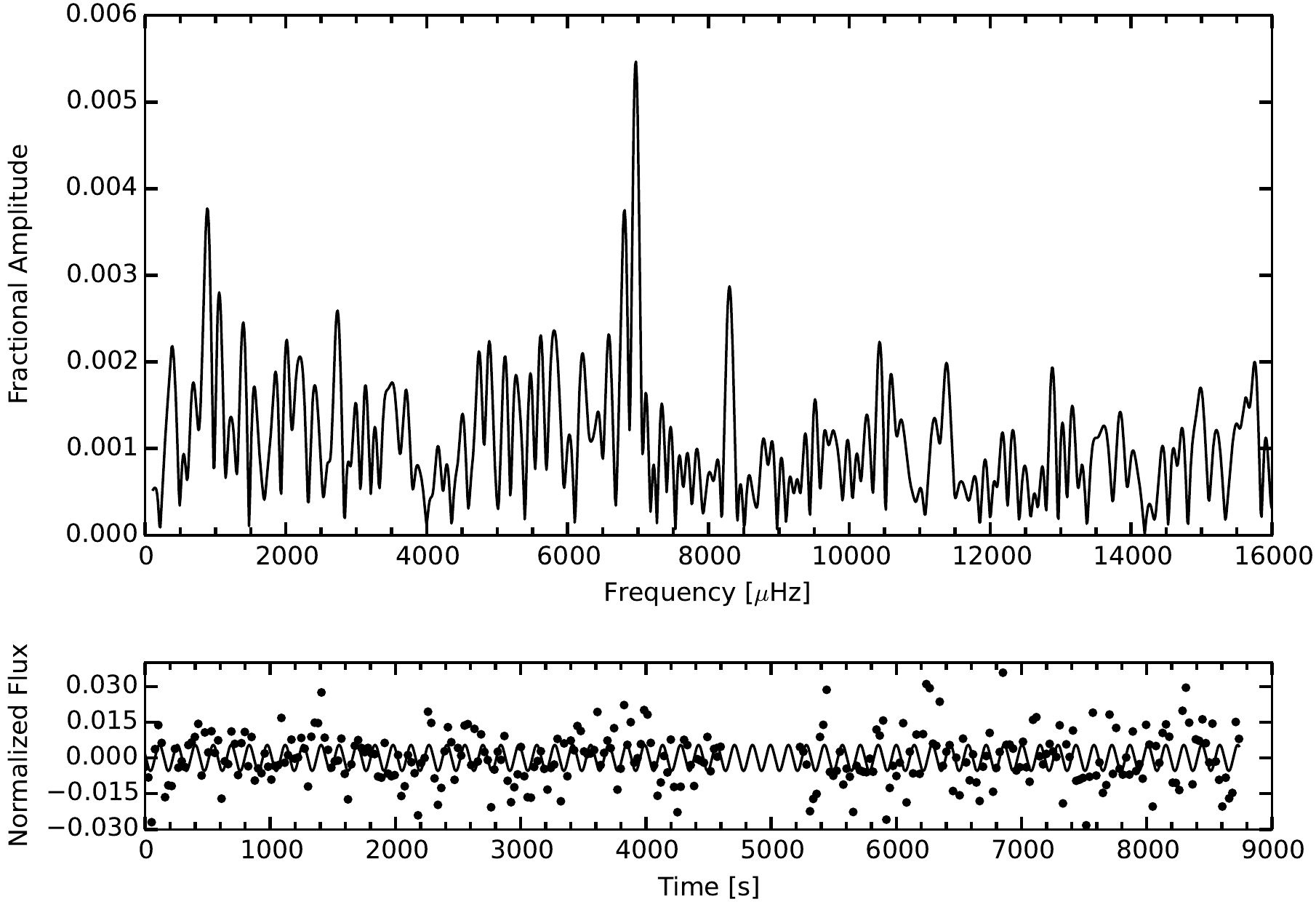}
    \caption{Optical time--series photometry of LAMOST J082517.99+113106.3 obtained with the robotic Skynet telescopes. The Lomb-Scargle periodogram (top panel) reveals the presence of a $\sim$6 ppt signal in the light curve (bottom) with frequency consistent with that found in the GALEX NUV data.}
    \label{Lamost_SKYNET}
\end{figure}

All data were bias--subtracted, flat--fielded, and dark--subtracted using standard procedures via the Skynet pipeline. We performed aperture photometry on LAMOST J082517.99$+$113106.3 using an in-house Python script. We chose the appropriate aperture radius to maximize S/N and used annuli to subtract sky brightness counts. Additionally, we tracked a nearby constant comparison star and ran the same aperture photometry procedure on it to remove atmospheric variations over the observing run. As with the GALEX observations, we calculated the LSP to look for any optical variations in the light curve. Figure \ref{Lamost_SKYNET} shows the resulting light curve and its amplitude spectrum. Our ground--based optical light curve reveals a photometric variation near the same frequency detected in the GALEX data. From least--squares fits of sine waves to the data, we report a ``white light'' amplitude of 5.4 $\pm$ 0.8 ppt with frequency 6971 $\pm$ 8 $\mu$Hz (period of 143.45 $\pm$ 0.18 s). 

Initially, we were a bit surprised at the relatively low optical amplitude of LAMOST J082517.99+113106.3. Other known sdBVs we observed had amplitudes in the optical that were approximately half that in the NUV. If LAMOST J082517.99+113106.3 followed the same trend, we would expect an amplitude of $\sim$10 ppt in the ground-based optical data. Further inspection of our Skynet images revealed that LAMOST J082517.99+113106.3 had an unresolved visual companion whose PSF overlapped heavily with that of the sdBV in the PROMPT-3 frames (which have a pixel scale of 1.4\arcsec\ per pixel). Consequently, the apertures we used when extracting photometry were heavily polluted by the companion, and our reported measurement for the optical amplitude must be underestimated. The visual companion is resolved in the Sloan Digital Sky Survey (SDSS J082518.17+113106.1) and sits $\sim$2.7\arcsec\ to the East \citep{sdss7}; it has SDSS colors signficantly redder than the sdB, consistent with a late F-type or early G-type star. Such a cooler companion should be approximately 5--6 mag fainter than the sdB in the NUV (see Figure 1 of \citealt{wad09}). In this case, our NUV pulsation amplitude should be unaffected even though the pair is unresolved in GALEX. In the Skynet optical images, we used SAOImage ds9 to estimate the flux ratio and find that the companion is approximately 30\% fainter than the sdB in the Clear filter. As such, a correction factor of $\sim$1.7 should be applied to our measured pulsation amplitude, which brings the true value closer to 9 or 10 ppt, much more consistent with the 2--to--1 ratio observed for the sdBV stars in Section \ref{OTFP}.  LAMOST J082517.99+113106.3 requires higher spatial resolution follow-up in order to accurately determine a precise optical pulsation amplitude.

\section{Discussion}
GALEX provides an enticing dataset to study UV--bright objects, and gPhoton makes such a study significantly easier. However, the GALEX dataset is not a golden ticket for those hoping to conduct a detailed and high-resolution study of variable objects. In our work with sdBs, we identified several pitfalls when working with GALEX data queried through gPhoton that future studies should be wary of (along with those discussed in \citealt {Mil16}). First, strong detrad signals near 8000$\mu$Hz (petal pattern fundamental), 16000 $\mu$Hz (petal pattern first harmonic), and below 1000 $\mu$Hz are present for many of the targets, even if they were not explicitly flagged as near the detector edge by gPhoton. Such instrumental signals can dominate the power in a periodogram, hiding lower--amplitude stellar pulsations in their window functions.  We used a pre--whitening technique to remove these signals so that we could look for lower--amplitude stellar pulsations, but if any true signals were present near the detrad frequencies, they were removed in the process, too.

\begin{figure}
	\centering
	\includegraphics[scale=0.33]{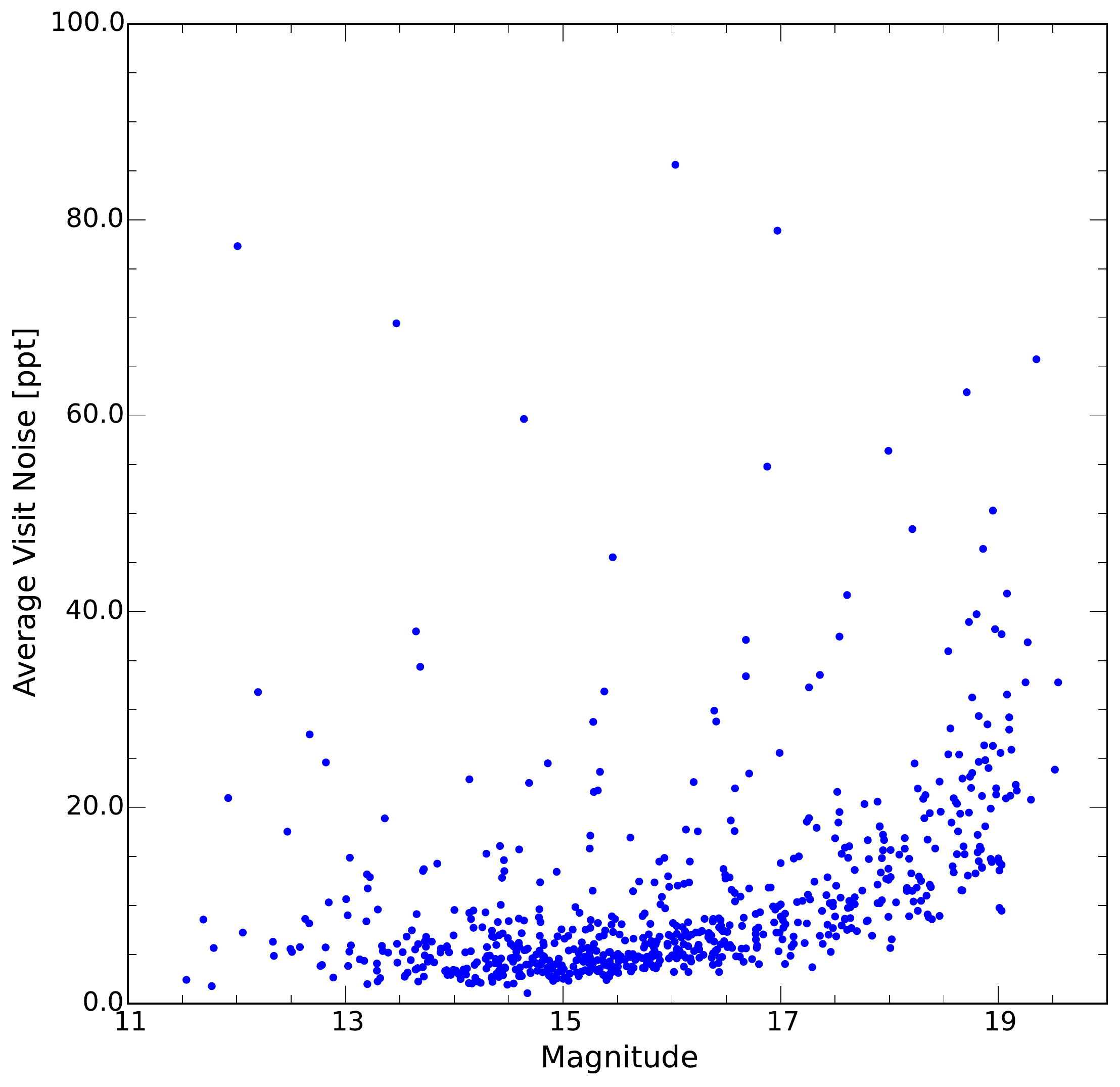}
    \caption{Average, single--visit LSP mean noise levels for all NUV light curves, plotted against the V magnitudes of the sdB targets. Most sdBV$_r$ stars have optical amplitudes at or below 10 ppt, an area heavily contaminated with noise.}
    \label{NoiseVsMag}
\end{figure}

GALEX's observational pattern and observing cadence give rise to other obstacles when studying sdBV$_r$ pulsations. The typical observing run length for a single spacecraft visit was relatively short, near 25 min. The most obvious downfall of such short visits is a poor signal--to--noise level in the data; low--amplitude signals ($<$10 ppt) are simply difficult to detect, even for bright objects. Figure \ref{NoiseVsMag} shows visit--by--visit LSP noise levels as a function of V magnitude for the majority of our sdB targets. Even relatively bright sdBs with V = 14-15 mag have a median single--visit noise level around $\sigma$ = 5 ppt. The 13 known pulsators extant within our dataset (Table \ref{KnownTab}) give a sense of these poor noise properties; we can see that for all 13 stars the mean noise levels are near or above their charectaristic UV pulsation amplitudes. If one were to apply a 4$\sigma$ or 5$\sigma$ criterion for the detection of new pulsation modes, most characteristic sdB pulsations would fall below this cutoff, masked by the noise. Another consequence of the short visit length is a less--than--desirable frequency resolution of 667 $\mu$Hz.   While 20-25 min can be sufficient to observe at least a few cycles of even the slowest sdBV$_r$ pulsation modes, a problem arises when multiple modes are present: they easily blend together in a power spectrum, as we observed for HS 2201+2610.

Improvements to the signal--to--noise ratio and frequency resolution can be achieved through multiple visits to the same target.  Unfortunately, few of our targets with detected pulsations had more than one visit.  Moreover, for those that did, GALEX's observing pattern gives rise to large gaps between visits, sometimes on the order of years. For this reason, it is nearly impossible to combine multiple GALEX visits together for a target when computing the LSP. We avoided this problem by breaking up data for each target by visit and computing an LSP for each visit individually; however this had the downfall of being computationally expensive, and complicating identification of pulsators as a signal would sometimes be present in some but not all visits. A few other methods for handling the breaks in data were initially considered, such as cross-correlating LSPs, or averaging LSPs together; however, due to counting and noise issues and these were rejected. 

In light of the above discussion points, we consider the GALEX survey an adequate tool for {\em identifying} pulsation modes in sdBV$_r$ stars in the NUV, but not {\em characterizing} them in detail.  Large--amplitude, single--mode pulsators are an exception to this, as they are immune to GALEX's poor single--visit frequency resolution and high noise levels.

\section{Conclusion}
For the majority of GALEX's lifespan, calibrated data from the spacecraft could only be used for single--frame NUV and FUV photometric analysis; with the recent development of gPhoton this is no longer a limitation. The massive catalog of GALEX data can now be used to extract time--series photometry on much of the sky. We use this newfound source of data to search for short--period UV variations in all hot subdwarf stars that were observed by GALEX.  While the observing cadence, visit lengths, and noise properties are less than ideal for observing and characterizing sdBV$_r$ puslations, we do detect UV pulsations in four previously--identified sdBV$_r$ stars and report their NUV amplitudes and frequencies. Some of our sdB targets not previously observed to vary show potential signals at the 4-$\sigma$ level or above and demand optical follow--up from the ground for confirmation.  We used the robotic Skynet telescope system to obtain optical photometry of one of these candidates, LAMOST J082517.99+113106.3, and confirm its nature as a new pulsating hot subdwarf star.  

The essential takeaway of our study is as follows: time--series aperture photometry can be extracted from GALEX data, but sdBs, despite being UV--bright objects, are not the most ideal candidates for study with this instrument. This is due to a number of factors, foremost among them that sdB pulsation frequencies often exist very close to the dither pattern frequency of GALEX, and sdB pulsation amplitudes are very near to the average noise level of GALEX.

\section*{Acknowledgements}
This research has made use of NASA's Astrophysical Data System. We would like to thank the Space Telescope Science Institute (STScI) and High Point University (HPU) for facilitating this research. Some/all of the data presented in this paper were obtained from the Mikulski Archive for Space Telescopes (MAST). STScI is operated by the Association of Universities for Research in Astronomy, Inc., under NASA contract NAS5-26555. Support for MAST for non-HST data is provided by the NASA Office of Space Science via grant NNX09AF08G and by other grants and contracts.

\software{gPhoton (Million et al. 2016), FaRVaE (https://github.com/tboudreaux/FaRVaE), SExtractor (Bertin \& Arnouts 1996), Scipy (Jones et al. 2001), DS9}
    

\end{document}